\documentclass{article}

     \usepackage[preprint]{neurips_2023}

\usepackage[utf8]{inputenc} 
\usepackage[T1]{fontenc}    
\usepackage{hyperref}       
\usepackage{url}            
\usepackage{booktabs}       
\usepackage{amsfonts}       
\usepackage{nicefrac}       
\usepackage{microtype}      
\usepackage{xcolor}         

\usepackage{amsmath}
\usepackage{amssymb}
\DeclareMathOperator*{\argmax}{arg\,max}
\DeclareMathOperator*{\argmin}{arg\,min}
\DeclareMathOperator{\E}{\mathbb{E}}
\usepackage{mhchem} 
\DeclareUnicodeCharacter{2212}{\textendash}
\usepackage{graphicx}

\title{Probabilistic Neural Transfer Function Estimation with Bayesian System Identification}

\author{%
Nan Wu \\
Saarland University \\
\And
Isabel Valera \\
Saarland University \\
\And
Fabian Sinz \\
University of Tübingen \\
University Göttingen \\
\And
Alexander Ecker \\
University Göttingen \\
\And
Thomas Euler \\
University of Tübingen \\
\And
Yongrong Qiu \thanks{Correspondence: qyrixpress@gmail.com}\\
University of Tübingen \\
University Göttingen \\
}

\begin{document}

\maketitle

\begin{abstract}
Neural population responses in sensory systems are driven by external physical stimuli. This stimulus-response relationship is typically characterized by receptive fields, which have been estimated by \textit{neural system identification} approaches. Such models usually requires a large amount of training data, yet, the recording time for animal experiments is limited, giving rise to epistemic uncertainty for the learned neural transfer functions. While deep neural network models have demonstrated excellent power on neural prediction, they usually do not provide the uncertainty of the resulting neural representations and derived statistics, such as the stimuli driving neurons optimally, from \textit{in silico} experiments. Here, we present a Bayesian system identification approach to predict neural responses to visual stimuli, and explore whether explicitly modeling network weight variability can be beneficial for identifying neural response properties. To this end, we use variational inference to estimate the posterior distribution of each model weight given the training data. Tests with different neural datasets demonstrate that this method can achieve higher or comparable performance on neural prediction, with a much higher data efficiency compared to Monte Carlo dropout methods and traditional models using point estimates of the model parameters. At the same time, our variational method allows to estimate the uncertainty of stimulus-response function, which we have found to be negatively correlated with the predictive performance and may serve to evaluate models. Furthermore, our approach enables to identify response properties with credible intervals and perform statistical test for the learned neural features, which avoid the idiosyncrasy of a single model. Finally, \textit{in silico} experiments show that our model generates stimuli driving neuronal activity significantly better than traditional models, particularly in the limited-data regime. Together, we provide a probabilistic approach for identifying neuronal representation with full distribution, which may help uncover the underpinning of high-dimensional biological computation.
\\
\end{abstract}

\section{Introduction}
Current neural interfaces allow to simultaneously record large populations of neural activity. 
In sensory neuroscience, such ensemble responses are driven by external physical stimuli (e.g., natural images), and their relation has been characterized by tuning curves or receptive fields (RFs; \citet{hubel1959receptive}). Such stimulus-response functions have been estimated by \textit{neural system identification} methods \citep[reviewed in][]{wu2006complete}. Classically,  they used a linear-nonlinear-Poisson (LNP) model or variants of it~\citep{chichilnisky2001simple,pillow2008spatio,huang2021estimating,karamanlis2021nonlinear} to predict responses to unseen stimuli such as white noise and natural images \citep{rust2005praise,qiu2021natural}. More recently, deep neural networks (DNNs) with multiple layers of non-linear processing have shown great success for learning neural transfer functions along the ventral visual stages from retina \citep{qiu2023efficient,mcintosh2016deep,batty2016multilayer} and primary visual cortex \citep{antolik2016model,klindt2017neural,ecker2018rotation,lurz2021generalization} to higher visual areas \citep{yamins2014performance,gucclu2015deep}. Moreover, through \textit{in silico} experiments, these models are able to generate specific stimulus to control neural activity and identify novel neuronal properties from a high-dimensional space \citep{bashivan2019neural,ponce2019evolving,walker2019inception,franke2021behavioral,hoefling2022chromatic}. For example, closed-loop paradigms show that performing gradient ascent on a deep model can yield most exciting inputs (MEIs) to drive a neuron's activity optimally~\citep{walker2019inception,tong2023feature}. 

Yet, these system identification approaches demand significant amounts of stimulus-response pair data for the model training, given the high dimensional stimulus space and the non-linear neural transformations \citep{qiu2023efficient,lurz2021generalization,cotton2020factorized}. Due to limited recording time for each experiment, the amount of data for fitting these models is restricted introducing epistemic uncertainty about the learned stimulus-response function. To estimate this uncertainty, traditional LNP methods obtain full posterior distribution of model parameters by leveraging a Bayesian framework to provide confidence intervals for the estimated RFs \citep{gerwinn2007bayesian,gerwinn2010bayesian,park2011bayesian,huang2021estimating}. However, DNN models rarely consider the uncertainty of the neuronal properties that are recovered from \textit{in silico} experiments. 

\begin{figure}
\begin{center}
   \includegraphics[width=0.7\linewidth]{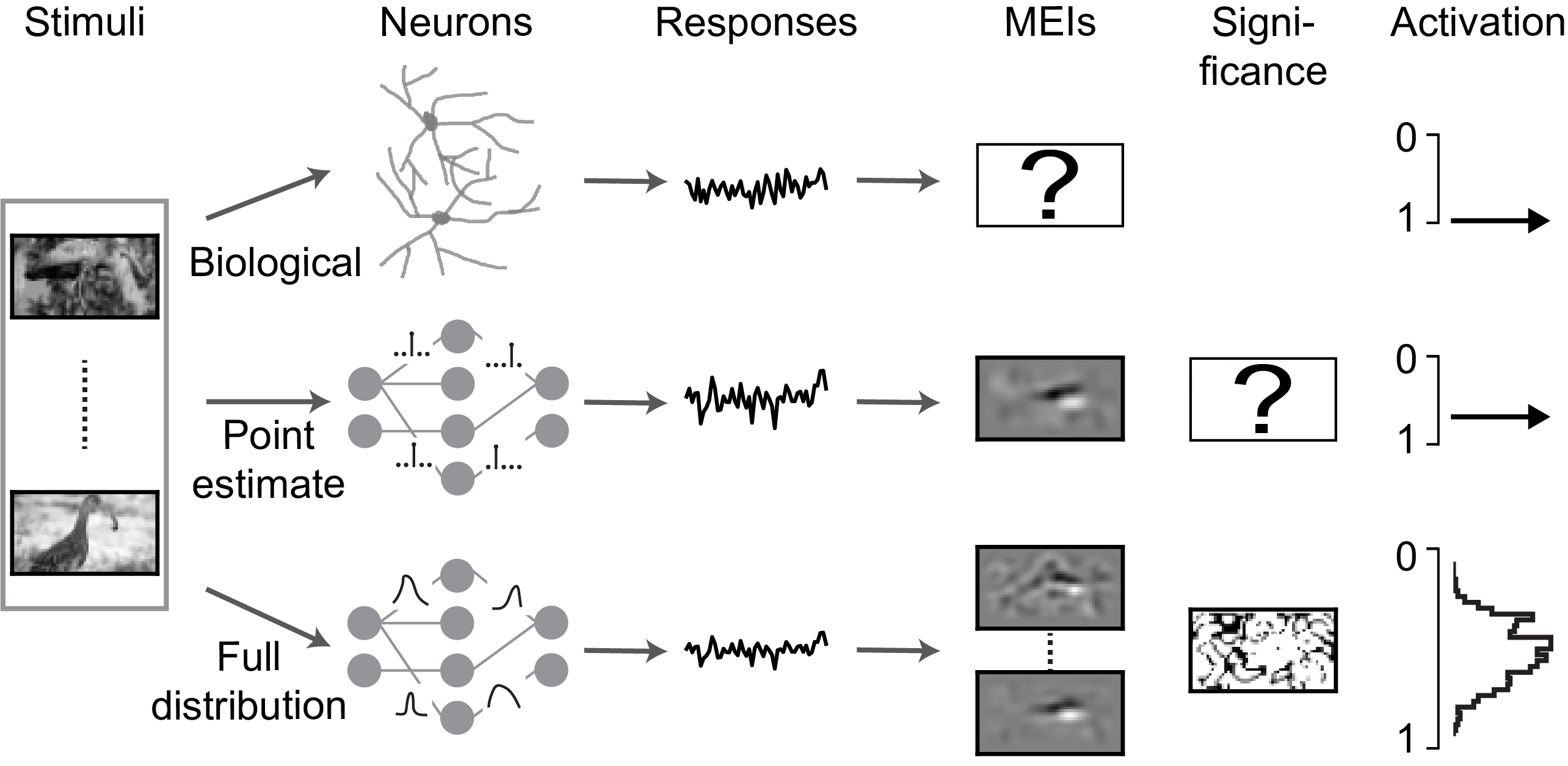}
\end{center}   
   \caption{
   Schematic of neural system identification for predicting responses. Biological neurons (top row; second column) respond to visual stimuli (first column) distinctly (third column), with an unknown MEI (fourth column) driving a cell with optimal activation (sixth column). Traditional system identification methods (center row) learn stimulus-response function and yield a MEI with unknown statistics (fitth column). Bayesian approaches (bottom row) learn distributions of model parameters to predict neuronal responses, yielding an infinite MEIs, whose significance map can be computed by sampling from posterior, to drive a neuron with credible intervals. 
   }
\label{fig:Fig1}
\end{figure}

Here, we propose a Bayesian system identification approach to estimate response features of neurons with uncertainties (Figure~\ref{fig:Fig1}). We test whether incorporating uncertainties by learning the full distribution of model parameters is beneficial for learning neural representations. To this end, we build a DNN model to predict responses to unseen visual stimuli by using variational inference to estimate the distribution of network weights, i.e., Bayes by Backprop \citep{hinton1993keeping,neal1998view,jaakkola2000bayesian,blundell2015weight}.

Our contributions are: (1) We incorporate weight variability in deep neural networks for identifying neural response functions with uncertainty and extend the Bayes by Backprop with a hyperparameter which effectively adjusts the sparsity of model parameters. (2) We apply our Bayesian models on different experimental datasets and find that our method can achieve higher or comparable performance on neural prediction, with a much better data efficiency, compared to Monte Carlo dropout methods and traditional models using point estimates of the model parameters. (3) Our approach with full posterior allows to estimate neural features with credible intervals and run statistical test for the derived MEIs, bypassing the idiosyncrasy of a single model. (4) Finally, simulation experiments demonstrate that the variational model yields stimuli that drive neuronal activation better than the traditional models, especially in the condition of limited training data. This supports that weight uncertainty, as implemented in our model, may contribute to a more efficient identification of non-linear neuronal response functions.

\section{Materials and methods}

\subsection{Models}
\paragraph{Variational model}
DNN for system identification can be seen as a probabilistic model: given the training data $ \mathcal{D} = (\textbf{x}_i,\textbf{y}_i)_{i} $ where $ \textbf{x}_i $ is an input (such as natural images) and $ \textbf{y}_i $ is the output (such as neural responses), we aim to learn the weights $ \textbf{w} $ of a network which can predict the output for the unseen stimuli (Figure~\ref{fig:Fig1}). Compared to a traditional method using point estimates of the weights, Bayesian approaches learn full distributions of these $ \textbf{w} $. Estimating the full posterior distribution of the weights $ P(\textbf{w}|\mathcal{D}) $ given the training data is usually not feasible. An alternative is to approximate $ P(\textbf{w}|\mathcal{D}) $ by a new distribution $ q(\textbf{w}|\theta) $ whose parameters $ \theta $ are trained to minimize the distance between the proxy and the true posterior, which is called variational inference \citep{hinton1993keeping,neal1998view,jaakkola2000bayesian,blundell2015weight}. Usually we use Kullback-Leibler (KL) divergence as a measure of distance between two distributions:
\begin{align}
\theta^* &= \argmin_{\theta} \textbf{KL} [q(\textbf{w}|\theta) || P(\textbf{w}|\mathcal{D})] \\
 &= \argmin_{\theta} \textbf{KL} [q(\textbf{w}|\theta) || P(\textbf{w})] - \E_{q(\textbf{w}|\theta)} [\log P(\mathcal{D} | \textbf{w})]
\end{align}
The optimization function can be viewed as a trade-off between the distance between the variational posterior and the selected prior and the likelihood cost. We can view it as a constrained optimization problem as \citep{higgins2016beta}:
\begin{align}
\argmax_{\theta} \E_{q(\textbf{w}|\theta)} [\log P(\mathcal{D} | \textbf{w})] \;\;\;\; \text{subject to} \;\; \textbf{KL} [q(\textbf{w}|\theta) || P(\textbf{w})]<\epsilon
\end{align}
Here $ \epsilon $ represents the specific distance between the variational posterior and the prior. According to KKT conditions \citep{kuhn1951nonlinear} and non-negative properties of KL divergence, we get:
\begin{align}
\mathcal{F} &= \E_{q(\textbf{w}|\theta)} [\log P(\mathcal{D} | \textbf{w})] - \beta_{v} ( \textbf{KL} [ q(\textbf{w}|\theta) || P(\textbf{w})] - \epsilon ) \\
&\geq \E_{q(\textbf{w}|\theta)} [\log P(\mathcal{D} | \textbf{w})] - \beta_{v} \textbf{KL} [ q(\textbf{w}|\theta) || P(\textbf{w})]
\end{align}
where $ \beta_{v} $ is non-negative and represents a Lagrangian multiplier. So the final loss function for the model is:
\begin{align}
    \mathcal{L} &= \beta_{v} \textbf{KL} [ q(\textbf{w}|\theta) || P(\textbf{w})] - \E_{q(\textbf{w}|\theta)} [\log P(\mathcal{D} | \textbf{w})] \\
    &\approx \sum_{i=1}^{n} \beta_{v} ( \log q(\textbf{w}^{(i)}|\theta) - \log P(\textbf{w}^{(i)}) ) - \log P(\mathcal{D} | \textbf{w}^{(i)})
\end{align}
Eq. (7) is a result of Monte Carlo sampling $n$ instances $ \textbf{w}^{(i)} $ from $ q(\textbf{w}|\theta) $ because we can not calculate (6) directly. 

Here, we implemented convolutional neural networks (CNNs) for all experiments. For a CNN using variational inference on model weights (variational model), we picked independent Gaussian distributions for the variational posterior and a scale mixture of two Gaussians for the prior \citep{blundell2015weight}. The log posterior was defined as $ \log q(\textbf{w}|\theta) = \sum_{k=1} \log \mathcal{N} (w_k | \mu, \sigma^2) $
where $ w_k $ denotes $k$th weight of the neural network and $ (\mu, \sigma) $ are the posterior parameters $\theta$. To keep $ \sigma $ non-negative, we parameterised it using $ \sigma = \log(1+\exp{(\rho)}) $. We selected the log prior $ \log P(\textbf{w}) = \sum_{k=1} \log ( \pi \mathcal{N} (w_k | 0, \sigma_1^2) + (1-\pi) \mathcal{N} (w_k | 0, \sigma_2^2) ) $
where $\pi$ is a mixture component weight ($ 0 \leq \pi \leq 1$) \citep{blundell2015weight,fortuin2021bayesian}. This prior, compared to a single Gaussian distribution, encourages sparseness in learned kernels, reminiscent of neural representations in visual systems \citep{field1994goal,olshausen1996emergence,olshausen1999learning,stevenson2008bayesian}. The likelihood loss depends on the specific task of the network. For neural system identification, we use  Poisson loss $ -\log P(\mathcal{D} | \textbf{w}) = \sum_{l} \hat{\textbf{r}}_l - {\textbf{r}_l} \log \hat{\textbf{r}}_l $, 
where $l$, $ \hat{\textbf{r}}_l $ and $ {\textbf{r}_l} $ denote neuronal index, prediction responses and true responses, respectively. 

\paragraph{Baseline and control models}
We used a CNN without any regularization as a baseline model (Appendix 5.1) and used a CNN with L2 regularization in each convolutional layer and L1 regularization in fully connected layer (L2+L1) as a control model. We adopted an ensemble of L2+L1 models with different initialization seeds as a second control model, whose predicted responses are the average of five model outputs. To examine the contribution from weight uncertainties, we built a maximum a posteriori (MAP) model which contains prior and likelihood terms in Eq. (7) as loss functions. Additionally, as a fourth control, we adopted a CNN with Monte Carlo dropout for probabilistic prediction; it used the same dropout rate for each model layer and in both training and test stages \citep{srivastava2014dropout,gal2016dropout}. 

\subsection{Dataset}
We tested our method on two publicly available datasets. 

The first dataset contains calcium signals driven by static natural gray-scale images for neurons in primary visual cortex (V1) of mice \citep{antolik2016model}. 
We used 103 neurons from the first scan field, whose single-trial responses to 1,600 images for training models and 200 for tuning hyperparameters. Then we used the mean of response repeats to 50 test images for evaluating models. 

The second dataset comprises \ce{Ca^2+} responses to natural green/UV images (36x64 pixels) for neurons in mouse V1 \citep{franke2021behavioral}. 
We selected the natural stimuli that were presented in both UV and green channels and used the neurons whose quality index ($QI = \mathrm{Var} [ \mathrm{E} [C ]_r ]_t / \mathrm{E}[ \mathrm{Var}[C ]_t ]_r$, time samples $t$ and repetitions $r$, a response matrix $C$ with a shape of $t \times r$, $\mathrm{E} [X]_d$ and $\mathrm{Var} [X]_d$ denoting the mean and variance along the dimension $d$ of $X$, respectively) of 10-repeat test responses were larger than 0.3. In this way, we had 161 neurons from one scan field, whose single-trial responses to 4,000 images for training and 400 for validation. Then we used mean of response repeats to 79 test images for evaluation.

\subsection{Training and evaluation}
We trained all models with a learning rate of 0.0003 for a maximum of 200 epochs using the Adam optimizer \citep{kingma2013auto}. We computed linear correlation (correlation coefficient, CC) between predicted and recorded responses, which was used to evaluate models on validation or test data. We tuned model hyperparameters and selected the ones as well as the respective epoch number with the best predictive performance on validation data. To keep the comparison fair, the test models shared similar network architecture for each dataset, except that the dropout model featured dropout layers.  


For each trained model, we estimated MEIs of all neurons by running gradient ascent on a random input image for 100 steps with a learning rate of 10 and we picked the stimulus with the highest activity \citep{erhan2009visualizing,walker2019inception}. All generated MEIs had the same mean and standard deviations as the training images. For the two probabilistic (variational and dropout) models, we ran the estimation for 100 times with Monte Carlo sampling, hence, we got 100 MEIs (matrix $ C $) for each recorded neuron. Note that we fixed the random seed/state for each sampling, in this way, model weights did not change stochastically during the iterative generation of each MEI. We defined MEI variance of one neuron as $ \mathrm{MEI \; variance} = \mathrm{E} [ \mathrm{Var} [C]_s ]_{hw} $ (sampling times $s$, stimulus height $h$, stimulus width $w$, and $ C $ with a shape of $s \times h \times w$). The overall MEI variance for a model was an average of MEI variances for the recorded neurons. 

In \textit{in silico} experiments, to measure the activation distribution of MEIs yielded from variational models for a neuron, we estimated 100 MEIs by sampling and one mean MEI by using the weight mean $ \mu $ from each seed. So we had 505 MEIs for five random seeds with one additional MEI which was the mean of the five mean MEIs, in total 506 MEIs. For L2+L1 models, we estimated five MEIs from different random seeds and also got one by averaging across these MEIs, in total 6 MEIs.

\section{Results}

\subsection{\texorpdfstring{$\beta_{v}$}{BETAV} balances model capacity and data likelihood} 

\begin{figure}
\begin{center}
   \includegraphics[width=\linewidth]{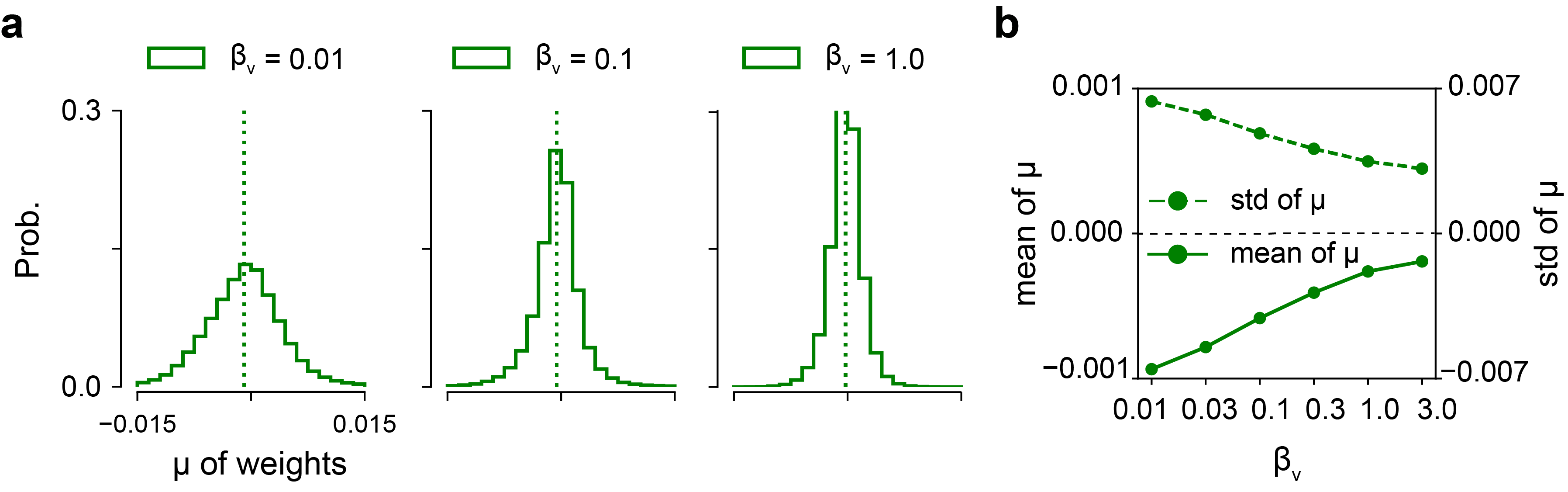}
\end{center}   
   \caption{Hyperparameter $\beta_{v}$ for regulating weight sparseness. 
   {\bf (a)} Distribution of the means ($ \mu $) of model weights for different $\beta_{v}$ values. Dotted lines indicate distribution means. 
   {\bf (b)} Mean and standard deviation for the distributions in (a).
   }
\label{fig:newFig2}
\end{figure}

Compared to a conventional evidence lower bound in Eq. (2), we used a Lagrangian multiplier $ \beta_{v} $ in (7) by borrowing the idea of constrained optimization from \citet{higgins2016beta}. In this way, Blundell and colleague's work can be seen as a special case of $\beta_{v}=1.0$ \citep{blundell2015weight}. We first analyzed the possible roles of $\beta_{v}$. We investigated it from the perspective of information theory, given that Eq. (7) has a similar form with the objective functions in deep variational information bottleneck \citep{alemi2016deep,tishby2000information} and $\beta$-VAE \citep{higgins2016beta,burgess2018understanding}.

The training objective jointly minimizes the KL divergence between the posterior $ q(\textbf{w}|\theta) $ and the prior $ P(\textbf{w}) $ and maximizes the data likelihood under the distribution $ q(\textbf{w}|\theta) $. The distribution distance becomes zero when $ q(\textbf{w}|\theta) = P(\textbf{w}) $. In the case of Gaussian posterior and mixture-of-Gaussians prior with mean zero (e.g., a distribution with $\pi=0.5, \sigma_1=1, \sigma_2=\exp{(-6)}$), the divergence decreases with the posterior mean moving close to zero and the posterior variance decreasing (as the exemplary prior has around 50\% of chance to be zero), which induces many zeros for weights $ \textbf{w} $ and increases the sparsity of model parameters. In the extreme case, all weights are equal to zeros and the model does not have any expressive power. In such case, the log likelihood $ \E_{q(\textbf{w}|\theta)} [\log P(\mathcal{D} | \textbf{w})] $ vanishes, indicating that the posterior $ q(\textbf{w}|\theta) $ is a bottleneck for maximizing the data likelihood. Therefore, $\beta_{v}$ can be interpreted as a coefficient to adjust model expressive power for fitting the data. 

Empirically, we examined the distribution of weight means ($ \mu $) for different $\beta_{v}$ values on the dataset 1 shared by by Antolik and colleagues \citep{antolik2016model}. Indeed, we found that with the increase of $\beta_{v}$, the mean of the distribution got close to zeros and the standard deviation decreased, indicating an increase of sparsity of model weights (Fig. \ref{fig:newFig2}). Similarly, we measured the percentage of the weights with absolute values below certain threshold ($0.005$) and also observed an increase of the ratio (data not shown). Therefore, the hyperparameter $\beta_{v}$ served to tune the model capacity via weight sparseness for data prediction.  

\subsection{System identification incorporates model uncertainty to predict neural responses}

\begin{figure}
\begin{center}
   \includegraphics[width=\linewidth]{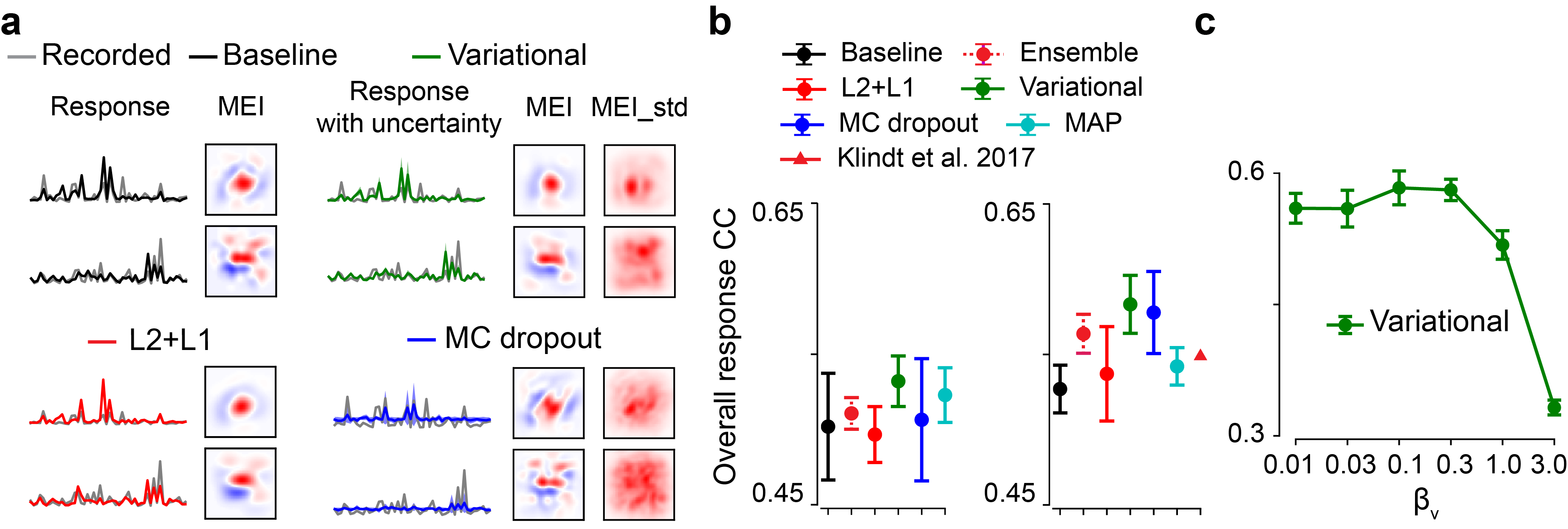}
\end{center}   
   \caption{ Neural prediction with weight uncertainty. 
   {\bf (a)} Mean recorded responses (gray) and predictive responses to natural stimuli(black, baseline; red, L2+L1; green, variational one with $\beta_{v}=0.1$; blue, MC dropout with dropout rate 70\%; shaded green and blue representing standard deviation for the variational and the dropout methods, respectively), estimated MEIs, as well as standard deviation of MEI (MEI\_std; only for two probabilistic models), for two exemplary neurons. MEI and MEI\_std use different color scales with red and blue indicating positive and negative values, respectively. Note that MEI has much larger absolute values than MEI\_std. 
   {\bf (b)} Predictive performance based on test data with different amounts of training data (left, 50\% of training data; right, 100\% of data) for 6 models (red dash, ensemble; cyan, MAP; 10 seeds per model), and the one used by \citep{klindt2017neural} (red triangle).
   {\bf (c)} Predictive model performance for different $\beta_{v}$ values.
   Error bars in (b) and (c) represent standard deviation of n=10 random seeds for each model. 
   }
\label{fig:newFig3}
\end{figure}

We trained the six models on the dataset 1 (Fig. \ref{fig:newFig3}a) and tuned their respective hyperparameters using validation data. For the variational model, we found the one with $\beta_{v}=0.1$ had best predictive performance with a sharp decrease when increasing $\beta_{v}$ till 1.0 or 3.0 (Appendix 5.2.1). We also observed that at training stage, the variational model presented a more stable performance on validation data compared to the baseline CNN, confirming the regularization effect of prior to prevent overfitting. 

Next, we selected the hyperparameters achieving the best performance on validation data for each model. To examine the feature properties learned by these models, we estimated the MEIs of recorded neurons and found that these models yielded antagonistic center-surround and Gabor filters in a local region, reminiscent of neural representations in early visual processing (\citep{hubel1959receptive,chichilnisky2001simple}; Fig. \ref{fig:newFig3}a). To compare the performance of neural prediction, we then evaluated all models using test data. For a probabilistic model, we ran model predictions for 100 sampling times and computed the mean and the standard deviation of neuronal responses. Interestingly, when we used the full training data, the variational and MC dropout models had similar predictive performance ($p=0.6159$, two-sided permutation test with n = 10,000 repeats). The variational one also outperformed the baseline, the L2+L1, the ensemble, the MAP ($p=0.0001$) and the model with shared feature space between neurons(\citep{klindt2017neural}; Fig. \ref{fig:newFig3}b). With half of training data, the variational method yielded a correlation slightly/non-significantly higher compared to the MC dropout method ($p=0.082$). The performance difference between variational and traditional methods using point estimates of parameters indicates the benefit of weight uncertainty for neural prediction. We then reanalyzed the influence of $\beta_{v}$ on prediction for the variational model using test data. Similar to the case with validation data, we noticed a rather steady predictive performance with increasing $\beta_{v}$ until a sudden drop at $\beta_{v}=1.0$ or 3.0, implying that a large Lagrangian multiplier imposing excessive sparsity on weights yields model underfitting. 

Together, the superior/equivalent performance of our variational approach suggests that incorporating weight uncertainty is beneficial for predicting neural responses.

\subsection{Probabilistic models learn variance of neural transfer functions} 

\begin{figure}
\begin{center}
   \includegraphics[width=\linewidth]{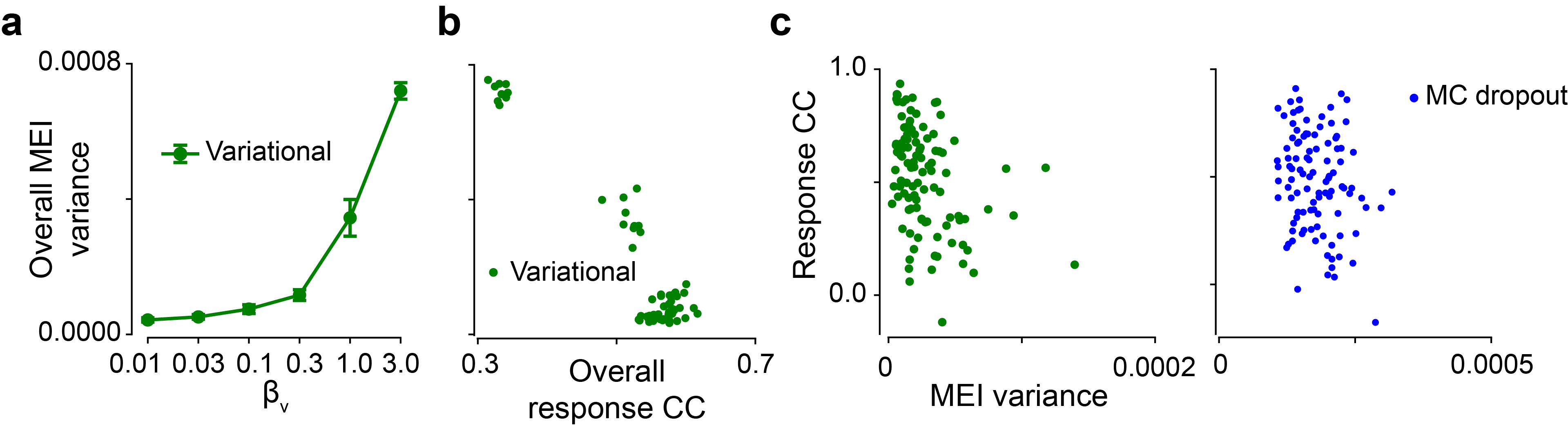}
\end{center}   
   \caption{ Neural transfer functions with variability. 
   {\bf (a)} Overall MEI variance for different $\beta_{v}$ values (10 seeds per model). 
   {\bf (b)} Scatter plot of overall response CC and overall MEI variance for 6 $\beta_{v}$ values and 10 seeds (each dot representing one model at each $\beta_{v}$ and each seed).
   {\bf (c)} Scatter plot of response CC and MEI variance for two probabilistic models at one random seed (each dot representing one neuron). 
   Error bars in (a) represent standard deviation of n=10 random seeds for each model. 
   }
\label{fig:newFig4}
\end{figure}

The variational and the MC dropout approaches enable us to learn stimulus-response functions with credible intervals. We next asked whether the variability of the learned transfer function was related to the predictive performance for the two probabilistic (variational and dropout) models. To this end, we measured the MEI variance for each neuron and the overall MEI variance for each model and relate them to the performance on predicting responses.  

We first investigated the influence of $\beta_{v}$ on the variability of the learned transfer functions for our variational model. Interestingly, we found a sudden increase of overall MEI variance at $\beta_{v}=1.0$ or $3.0$ (Fig. \ref{fig:newFig4}a), where an abrupt drop of model performance was present (cf. Fig. \ref{fig:newFig3}c). This opposite change between MEI variability and predictive performance was confirmed by the negative correlations between overall MEI variance and overall response CC ($CC=-0.95, p<0.0001$; Fig. \ref{fig:newFig4}b). 
Additionally, this negative correlation was also reflected at neuronal level. Both the variational and the MC dropout models had a negative correlation between response CC and MEI variance for the recorded neurons ($CC=-0.37, p=0.0001$ and $CC=-0.23, p=0.02$ for the variational and the MC dropout, respectively; Fig. \ref{fig:newFig4}c), indicating that, for a trained probabilistic model, neurons with higher predictive performance have higher confidence on its estimated MEI. 

In summary, these results demonstrate that a probabilistic model with smaller uncertainty on the learned stimulus-response function yields higher predictive performance.

\subsection{Variational model features high data efficiency on neural prediction} 

\begin{figure}
\begin{center}
   \includegraphics[width=0.8\linewidth]{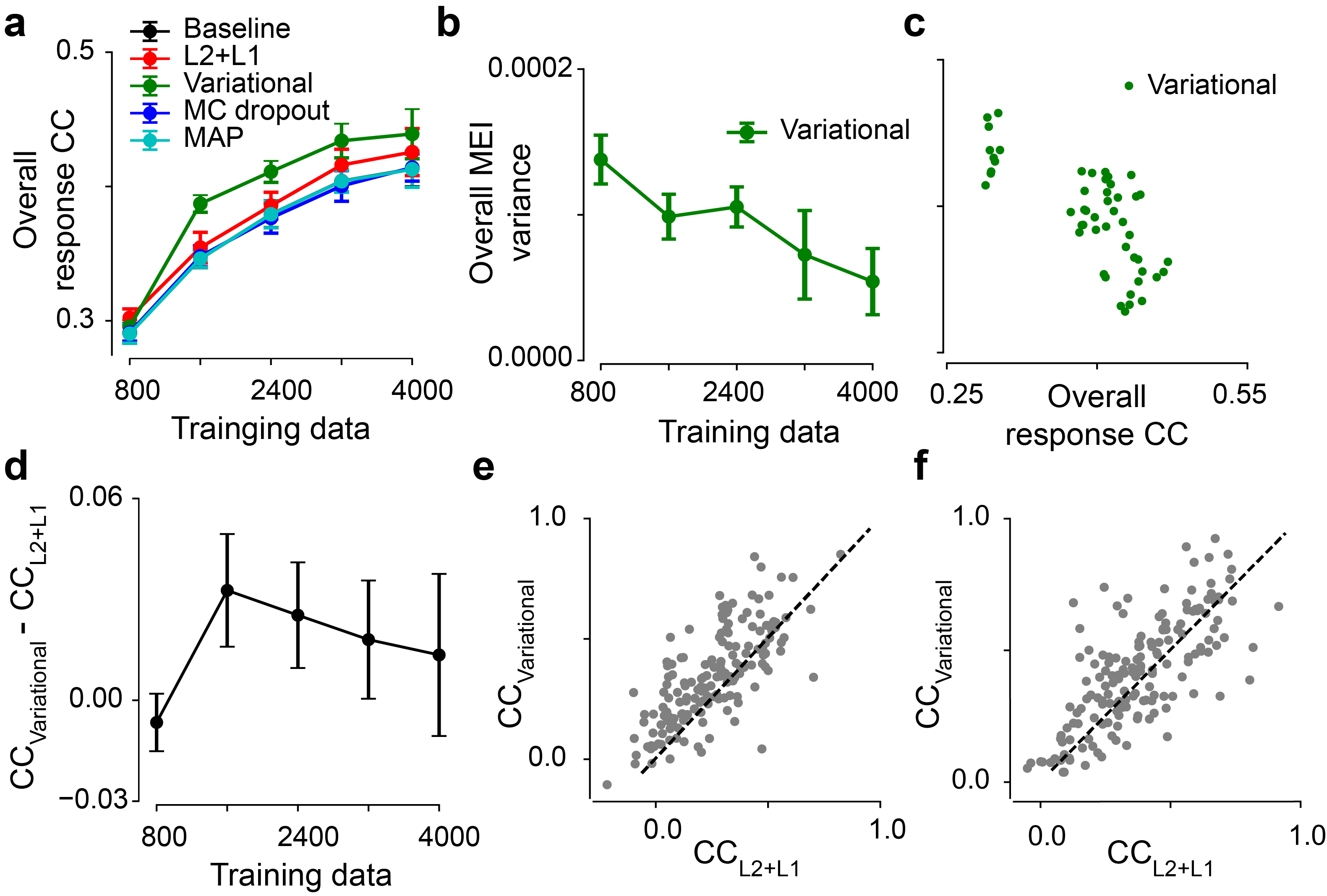}
\end{center}   
   \caption{ Variational models on the second dataset. 
   {\bf (a)} Model performance based on test data of the second dataset with different amounts of training data for five models (n=10 random seeds per model). 
   {\bf (b)} Overall MEI variance for different amounts of training data for variational models (10 seeds per model). 
   {\bf (c)} Scatter plot for overall response CC and overall MEI variance for different amounts of training data and at 10 seeds. Each dot representing one model. 
   {\bf (d)} Performance difference between the variational and the L2+L1 models.
   {\bf (e)} Scatter plot of model predictions for the variational model and the L2+L1 model at one random seed when using 40\% training data. Each dot representing one neuron.
   {\bf (f)} Like (e) but using 100\% training data. 
   Error bars in (a), (b) and (d) represent standard deviation of n=10 random seeds for each model. 
   }
\label{fig:newFig5}
\end{figure}

Here we applied our method on the second dataset shared by Franke and colleagues (\citep{franke2021behavioral}). After hyperparameter tuning, we selected $\beta_{v}=0.3$ for the variational network and evaluated the five models on test data. 

We first examined the relationship between the uncertainty of the learned stimulus-response function and the performance on predicting responses. We expect that, with more data used for training, the model yields better prediction along with smaller variance for the learned MEIs. We focused on the variational method. Indeed, when more training data was used, the predictive model performance increased (Fig. \ref{fig:newFig5}a) while the overall MEI variance decreased Fig. \ref{fig:newFig5}b, with a negative correlation between them ($CC=-0.73, p<0.0001$; Fig. \ref{fig:newFig5}c). Note that we did not observe a steady decrease of the overall response variance (Appendix 5.2.2). 

Next, we investigated whether the performance difference between the variation and the L2+L1 model was sensitive to the training data size (Fig. \ref{fig:newFig5}d). We observed that the variational method had higher correlations except for the case of extremely little data (20\%). The difference peaked at 40\% with an increase of 9\% ($p<0.0001$, two-sided permutation test with n = 10,000 repeats) and gradually decreased with more training data, indicating the benefit of variational inference for system identification. We also compared the predictive performance on individual neurons at one random seed, the Bayesian model outperformed the L2+L1 one for the conditions of 40\% ($p<0.0001$) and 100\% (non-significantly; $p=0.0927$) training data. 

Together, compared to a traditional method, our Bayesian approach with weight uncertainty yielded higher predictive performance with a higher data efficiency.

\subsection{Variational model yields stimuli driving high neuronal activation}

\begin{figure}
\begin{center}
   \includegraphics[width=\linewidth]{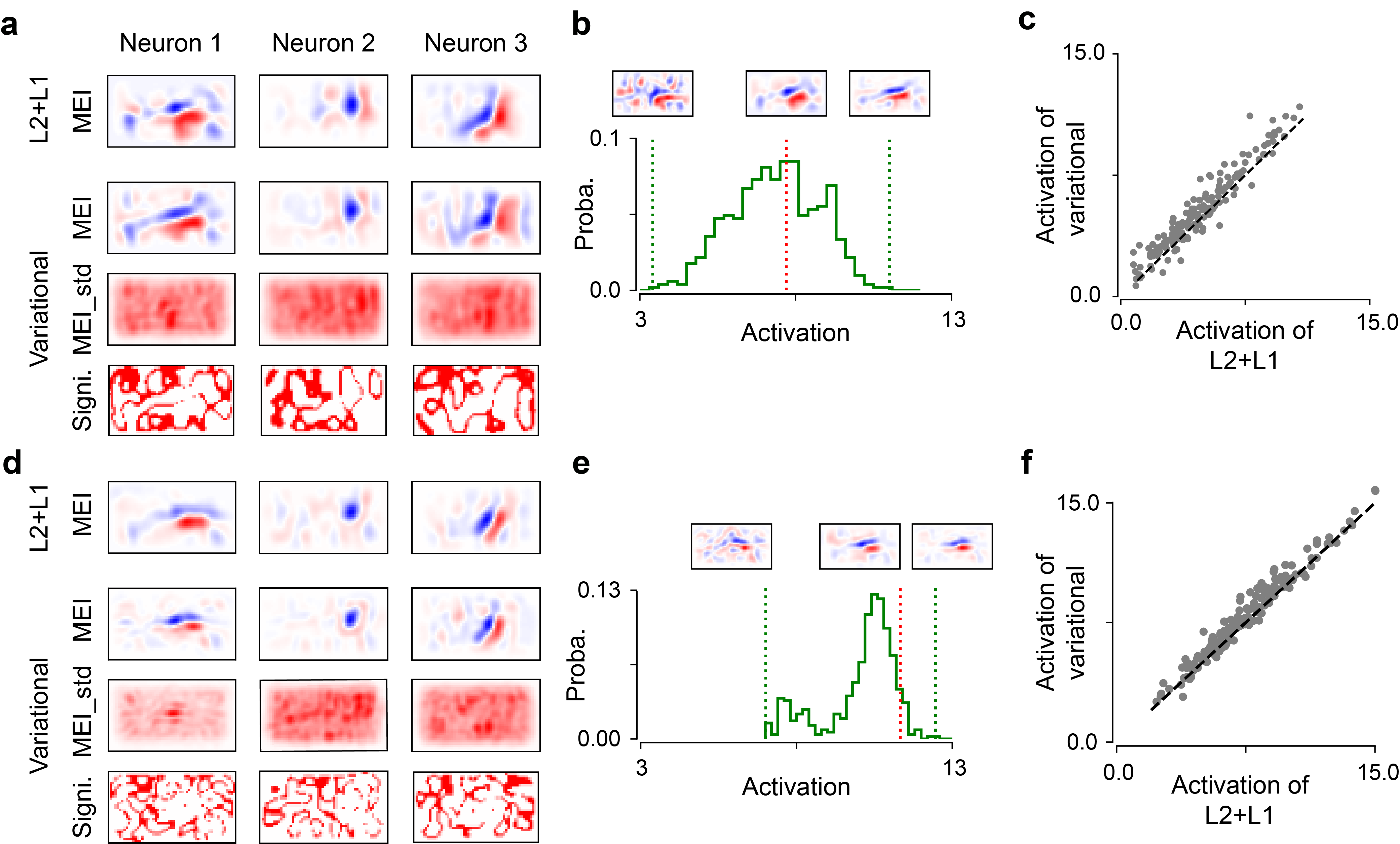}
\end{center}   
   \caption{ \textit{In silico} experiments of neuronal activity with derived MEIs. 
   {\bf (a)} Estimated MEIs for L2+L1 (first row) and variational (second row) models, MEI\_std (third row), as well as significance map (fourth row; white, $p<0.01$, one-sample two-sided permutation test against zero for 10,000 repeats), for three exemplary neurons when using 40\% of training data. MEI and MEI\_std in the UV channel, with different color scales. Note that MEI has much larger absolute values than MEI\_std.
   {\bf (b)} 1D histogram of neuronal activity driven by the generated MEIs from the variational model for Neuron 1 when using 40\% of training data. Insets: example MEIs with corresponding activation indicated by dotted lines (red, maximum of L2+L1; green, variational).
   {\bf (c)} Scatter plot of activation driven by MEIs yielded from variational (using the weight mean $\mu $) and L2+L1 models at one random seed when using 40\% of training data. Each dot representing one cell. 
   {\bf (d,e,f)} Same with (a), (b) and (c), but using 100\% of training data.
   }
\label{fig:newFig6}
\end{figure}

Bayesian methods with full posterior provide an infinite ensemble of models for computing MEIs and allow to perform statistical tests for the derived features. We focused on the model using variational inference and the one using L2 and L1 regularization with 40\% and 100\% of training data. 
We found that these learned filters resembled neural features in the early visual system \citep{hubel1959receptive,chichilnisky2001simple} and localized more in the visual field with more training data (Fig. \ref{fig:newFig6}a,d). Like for the first dataset (Fig. \ref{fig:newFig3}a), MEI\_std was not uniform across visual space, e.g., some presented Gaussian or bar shapes. Additionally, we examined whether the posterior of each pixel differs significantly from zero for the 100 sampled MEIs and found that the significance map may indicate zero-crossings in visual representations. 

To examine the effectiveness of our variational approach, we fed the verified MEIs from \citet{franke2021behavioral} into the model trained by full data and estimated the neuronal activation. Indeed, we observed that each cell was driven most by its respective preferred stimulus (Appendix 5.2.4). 

To further assess which method generates the more exciting stimuli for each cell, we conducted \textit{in silico} experiments using a held-out L2+L1 model trained by full data as a digital testbed. We used a CNN model with regularization instead of other models as previous studies have demonstrated its feasibility on yielding cells' preferred stimuli \citep{walker2019inception,franke2021behavioral,hoefling2022chromatic,tong2023feature}. For an example neuron, we measured the responses for all the 506 MEIs yielded from five variational models, and observed that these stimuli drove this neuron with quite different activity, with the maximum response larger than the maximum one yielded (from 6 MEIs) by the traditional models (Fig. \ref{fig:newFig6}b,e). With more training data, the activation distribution shifted towards higher mean with smaller variance. 
Additionally, we compared the activation on individual neurons for two methods (Fig. \ref{fig:newFig6}c,f), and observed that the Bayesian approach yielded significantly higher responses for the condition using 40\% of training data ($p=0.0473$ and $p=0.2114$ for 40\% and 100\% of data, respectively; two-sided permutation test with n = 10,000 repeats). 

In summary, our variational model allowed statistical test for the derived response functions and yielded the stimuli driving neurons better than traditional methods, suggesting that weight uncertainty benefits the learning of neural representations.

\section{Discussion}
We presented a Bayesian approach for identification of neural properties by incorporating model uncertainty through learning the distribution of model weights, aiming to estimate neural features with credible intervals. Our empirical results on different datasets show that the variational method had higher or comparable predictive performance, especially in the limited data regime, compared to methods using dropout or traditional methods learning point estimates of model parameters. 
Moreover, by sampling from posterior distribution of model weights, our approach enabled to provide credible intervals and test statistics for the learned MEIs, avoiding the idiosyncrasy of a single model. Finally, \textit{in silico} experiments show that the variational model yielded the MEIs driving neurons with higher activity compared to the traditional model, in particular when limited data were used for training. This suggests that model uncertainty contributes to learning neural transfer functions with a high data efficiency. 

\subsection{Relation to trial-to-trial variability}
Neural information process is probabilistic, i.e., neurons respond with trial-to-trial fluctuations to a repeated presentation of a stimulus \citep{perkel1967neuronal,stein1967some}. Response variability is found across neural systems, originating from diverse factors, such as synapse variation, channel noise, brain state, and attention \citep{faisal2008noise,mitchell2009spatial,cohen2008context,cohen2009attention,ecker2010decorrelated,ecker2014state}. Additionally, the variability between populations of neurons are correlated. In a simplified case, a pair of neurons may present correlations for the single-trial responses, i.e., pairwise noise correlation, which also contributes to neural coding (\citep{abbott1999effect}; reviewed in \citep{averbeck2006neural,kohn2016correlations,doiron2016mechanics,da2021geometry}). Such response variability is inherent in neural data itself and is a kind of aleatoric, but not the epistemic uncertainty. We note that the standard deviations of the estimated MEIs from our models decreased with the increasing amounts of training data, suggesting that the variability of the sampled predicted responses may not be related to the response uncertainty in biological neurons or our models may predict a mix of both uncertainties (Appendix 5.2.3). 

We did not observe a negative correlation between the predictive performance and the variance of predicted responses (cf. MEI variance; Appendix 5.2.1, 5.2.2), which might be related to the differential response variability driven by distinct stimuli \citep{goris2014partitioning,ecker2014state}. This also suggests a future study to compare between the uncertainty of recorded and predicted responses and investigate the response variance from different models (such as variational inference vs. MC dropout). Additionally, it might be interesting to use proper scoring rules, e.g., Continuous Ranked Probability Score, to evaluate the quality of predictive uncertainty for the Bayesian models \citep{matheson1976scoring,gneiting2007strictly}.

\subsection{Necessity of uncertainty quantification for yielded preferred stimuli}

Though DNN approaches have demonstrated remarkable power in predicting neural responses to diverse stimuli and generating novel hypothesis about neuronal features, they require significant amounts of stimulus-response pair data for the training. Besides the epistemic uncertainty introduced by limited data, such hypothesis also entails further closed-loop animal experiments to verify the derived properties, which consumes much experimental time \citep{walker2019inception,tong2023feature}. Still, it is impossible to confirm the yielded preferred stimuli for all neurons across the high-dimensional stimulus space with experiments. Practically, only a subset of cells are selected for verification. Therefore, it is critical to quantify the uncertainty of the yielded representations for all recorded neurons \citep{richards2019deep,saxe2021if}. 
Additionally, the credible interval of the derived features offers an opportunity to generate an ensemble of infinite preferred stimuli. An interesting study would be to compare the neuronal activity driven by these similar MEIs in animal experiments, which may allow to test the robustness of the biological system. 

We note that, even for a model using point estimate of parameters (such as L2+L1), it may yield different preferred stimuli by initializing the MEI generation randomly. Yet, this uncertainty depends on the starting points of non-convex optimizations, rather than the training data. Empirically, we found that such variance was quite stable when using L2+L1 models with different amounts of training data and was also much smaller than the MEI variance we computed (cf. Fig. \ref{fig:newFig5}b). Therefore, the measure of epistemic uncertainty calls for a Bayesian framework or an ensemble of many models.

\subsection{Future work \& general impact}

Incorporating uncertainty to DNNs have flourished in recent years \citep[reviewed in][]{abdar2021review,gawlikowski2023survey}, including Bayesian methods which specify a prior distribution for network weights and approximate the full posterior given the training data using different tricks such as variational inference \citep{blundell2015weight,posch2020correlated}, Laplace approximation \citep{mackay1992bayesian,ritter2018scalable} and expectation propagation \citep{li2015stochastic}. Non-Bayesian methods include applying MC dropout in the network \citep{gal2016dropout} or training an ensemble of models that are initialized by different seeds \citep{lakshminarayanan2017simple}. While these methods are powerful to predict uncertainty, it would be interesting to investigate biologically inspired methods such as adding noise to network parameters/activation in the future. 
Specifically, our variational approach incorporating model uncertainty did not predict the trial-to-trial variability. Such response fluctuation depends on many conditions, including biochemical process, internal brain states and engaged behavioral tasks \citep{faisal2008noise,mitchell2009spatial,ecker2014state,goris2014partitioning}. These factors have been described by a low-dimensional latent state models \citep{yu2008gaussian,ecker2014state,bashiri2021flow}. Therefore, a potential extension of our method could be a variational network incorporated with latent state variables.

Our \textit{in silico} experiments indicate that the stimuli generated by the variation model driving higher neuronal activation than the CNN with regularization, which requires future animal experiments to test. Additionally, we noticed that the MEI\_std was not uniform in the visual field for each neuron and its location was not overlaid with the central MEI, for example, it seems to sit on the surround of the corresponding MEI. It would be interesting to examine and quantify the MEI uncertainty in regard of visual space, which might be related to contextual sensory processing \citep{hock1974contextual,chiao2003contextual,fu2023pattern}.

More generally, why do we care about the uncertainty of the estimated neural representations? Even with closed-loop experiments, it is impossible for us to test all potential (preferred) inputs for the recorded neurons \citep{walker2019inception,franke2021behavioral,tong2023feature}. Therefore, we always expect to have a confidence interval for the test statistics. Besides, a Bayesian model offers a manner to generate many stimulus candidates by sampling for stimulating neural systems, which may offer new insights for understanding the biological computation.

\section*{Acknowledgments}
We thank Philipp Berens, Katrin Franke, Suhas Shrinivasan and Ziwei Huang for helpful discussions. This work was supported by the German Research Foundation (DFG; SFB 1233, Robust Vision: Inference Principles and Neural Mechanisms, projects 10 and 12, project number 276693517; SFB 1456, Mathematics of Experiment, project number 432680300). The funders had no role in study design, data collection and analysis, decision to publish, or preparation of the manuscript. 

\section*{Declaration of Interests}
The authors declare no competing interests. 

\section*{Code Availability}
Code will be available upon publication.

\bibliographystyle{plainnat}
\bibliography{main.bib}

\newpage
\section{Appendix}
\subsection{Model details}
The CNN model for the first dataset shared by Antolik and colleagues consisted of a convolutional layer (24x1x9x9, output channels x input channels x image width x image height),  a rectified linear unit (ReLU) function, another convolutional layer (48x24x7x7, output channels x input channels x image width x image height), another ReLU function, and --- after flattening all dimensions --- one fully connected (FC) layer (103x13872, output channels x input channels), followed by an exponential function. We used stride=1 and no padding for both convolutional layers. 
We trained the four models and tuned their respective hyperparameters. For the variational one, we tested different parameters for prior distribution on validation data, such as $\pi=0$ or $\pi=0.5$, $\sigma_1=1$ or $\sigma_1=100$, $\sigma_2=\exp{(-3)}$ or $\sigma_2=\exp{(-6)}$, and found that a scale mixture of two Gaussians had similar predictive performance, higher than one Gaussian distribution. As the predictive performance was similar for distinct priors on model layers, we used the same prior distribution with parameters $\pi=0.5, \sigma_1=1, \sigma_2=\exp{(-6)}$ for all layers. We also examined the number of Monte Carlo sampling times for model training and found that the predictive performance was similar for different numbers. Therefore, we used 1 or 2 sampling times for all model training. 

The CNN model for the second dataset shared by Franke and colleagus contained a convolutional layer (48x2x9x9), a ReLU function, another convolutional layer (48x48x7x7), another ReLU function, and one FC layer (161x52800), followed by an exponential function. We used stride=1 and no padding for both convolutional layers. 

For each CNN model, we tested different numbers (1-5) of blocks (each consists of a convolutional layer and a ReLU layer) and different channel numbers (8, 16, 24, 32, 40 and 48) of each convolutional layer, and selected the ones which yielded (near) optimal predictive performance on validation data. We used small numbers when the performance was similar across models. For each dataset, the six methods used similar model architecture with the baseline CNN, except that the dropout model had dropout layers after the two ReLU functions and the FC layer. 

To estimate the uncertainty of neuronal responses for a trained probabilistic model, We ran the model prediction for 100 sampling times. We defined variance of predicted response for one neuron as $ \mathrm{Response \; variance} = \mathrm{E} [ \mathrm{Var} [D]_s ]_{n} $ (sampling times $s$, test stimulus number $n$, and response matrix $ D $ with a shape of $s \times n$). The overall response variance for a model was an average of response variances for the recorded neurons. To calculate the variance of recorded response for a neuron, we replaced the sampling times with the repeated times of the presented test stimulus.

\subsection{Additional results} 

\subsubsection{Neural prediction for first dataset} 


\begin{figure}[!htb]
\begin{center}
   \includegraphics[width=\linewidth]{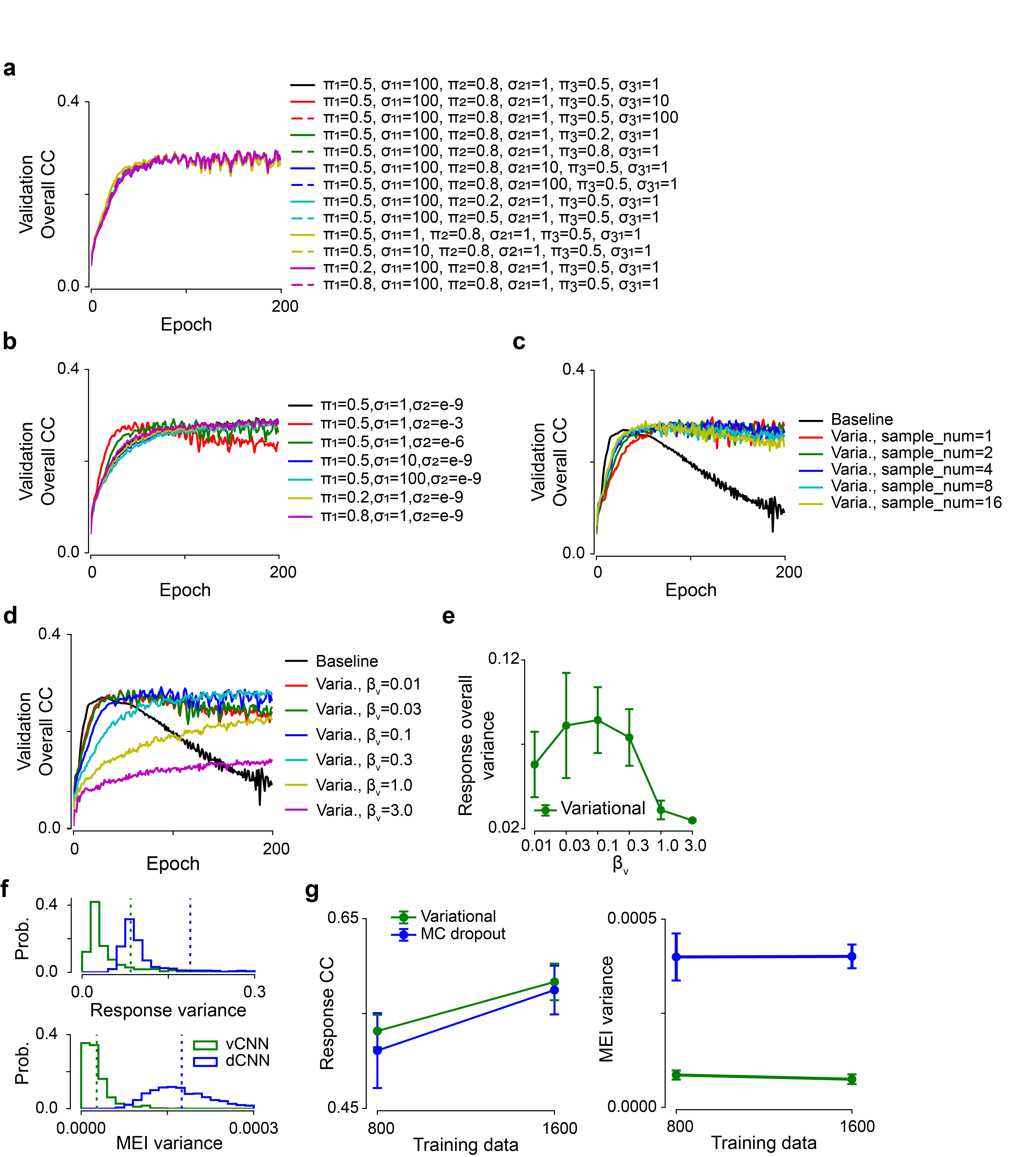}
\end{center}   
   \caption{Neural prediction for first dataset. 
   {\bf (a,b)} Predictive performance (correlation coefficient, CC) based on validation data during training for variational models ($\beta_{v}=0.1$) with different prior distributions. All layers adopted the same $\sigma_2=\exp{(-6)}$ with different $\pi$ and $\sigma_1$ values (a), or with the same parameters of prior distribution (b). We picked $\pi=0.5, \sigma_1=1, \sigma_2=\exp{(-6)}$ for subsequent model training. 
   {\bf (c)} Predictive performance based on validation data during model training for different numbers of Monte Carlo sampling. We picked number=1 or 2 to save training time. 
   {\bf (d)} Model performance based on validation data during training for the baseline and variatonal models with different $\beta_v$ values. 
   {\bf (e)} Overall variance of predicted responses to test stimuli for different $\beta_{v}$ values.
   {\bf (f)} Histogram of response variance (top) and MEI (RF) variance (bottom) for the variational and the MC dropout models. Dotted line represents the mean of histogram. 
   {\bf (g)} Model performance (left) based on test data and RF overall variance (right) for two probabilistic models with different amounts of training data. 
   Error bars in (e) and (g) represent standard deviation of n=10 random seeds for each model. 
   }
\label{fig:Fig2_S1}
\end{figure}

\newpage

\subsubsection{Neural prediction for second dataset} 

\begin{figure}[!htb]
\begin{center}
   \includegraphics[width=\linewidth]{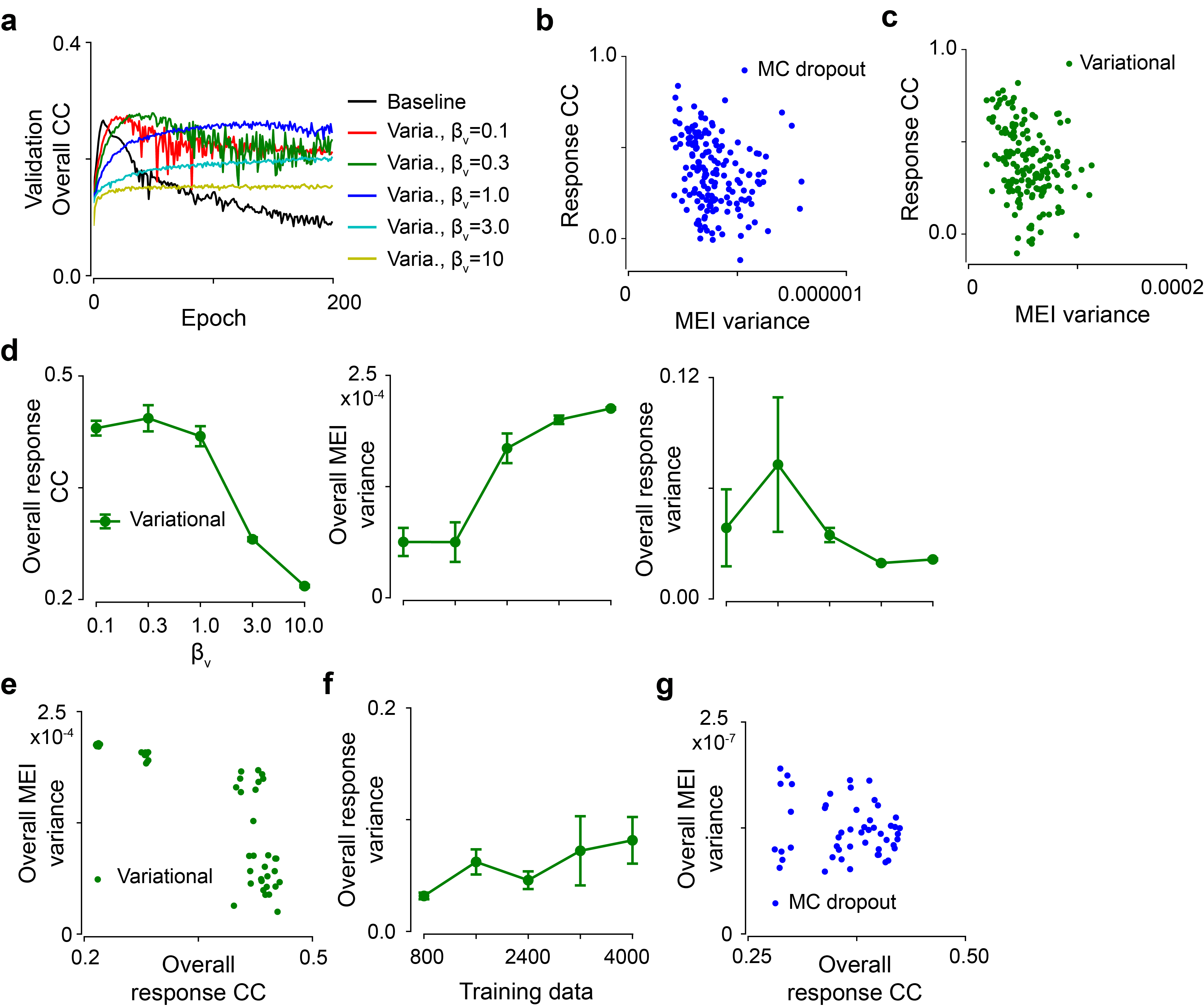}
\end{center}   
   \caption{Neural prediction for second dataset. 
   {\bf (a)} Model performance based on validation data during training for the baseline and the variational models with different $\beta_v$ values. 
   {\bf (b,c)} Scatter plot of response CC and MEI (RF) variance for MC dropout (b) and variational (c) models for 10 seeds ($CC=-0.25, p=0.001$ and $CC=-0.34, p<0.0001$ for dropout and variational one, each dot representing one neuron at one random seed). 
   {\bf (d)} Predictive performance, overall RF variance and overall response variance for variational models with different $\beta_{v}$ values. 
   {\bf (c)} Predictive performance based on validation data during model training for different numbers of Monte Carlo sampling. We picked number=1 or 2 to save training time. 
   {\bf (d)} Model performance based on validation data during training for the baseline and the variational ones with different $\beta_v$ values. 
   {\bf (e)} Scatter plot for overall response CC and overall RF variance for the variational methods with different $\beta_v$ values (d) and at 10 seeds ($CC=-0.82,p<0.0001$). Each dot represents one model. 
   {\bf (f)} Overall response variance for different amounts of training data for the variational models (10 seeds per model). 
   {\bf (g)} Scatter plot for overall response CC and overall RF variance for the dropout model with different amounts of training data and at 10 seeds ($CC=-0.17,p=0.24$). Each dot represents one model. 
   Error bars in (d) and (f) represent standard deviation of n=10 random seeds for each model. 
   }
\label{fig:Fig4_S1}
\end{figure}

\newpage

\subsubsection{Variance of predicted vs. recorded responses for second dataset} 

\begin{figure}[!htb]
\begin{center}
   \includegraphics[width=0.8\linewidth]{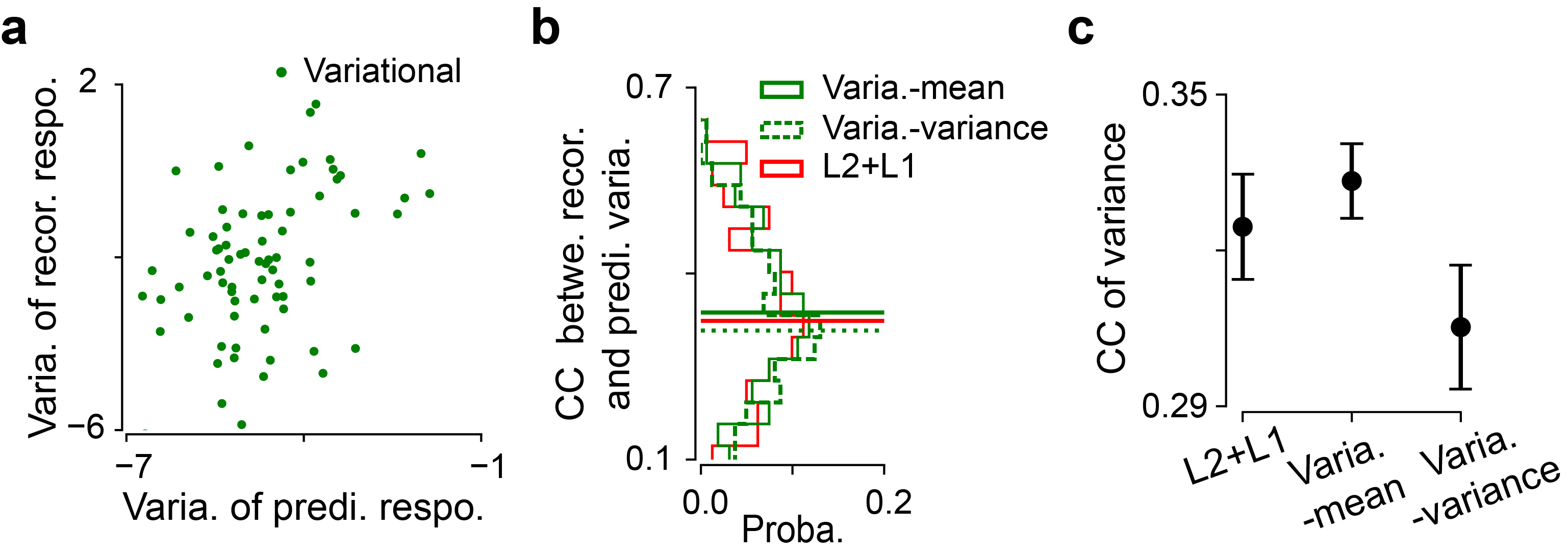}
\end{center}   
   \caption{Variance of predicted vs. recorded responses for second dataset. 
   Using the trained models, we tested whether the variance of predicted responses was related to the variance of recorded responses for each neuron. We first estimated the predicted response variance to a stimulus. For the L2+L1 model, as the mean of neural responses is proportional to the variance, we used the model output (a single predicted value) as a substitute. For the variational one, we either used the mean of predicted responses (multiple sampling times) as a substitute or calculated the response variance explicitly.     
   {\bf (a)} Scatter plot (axes in log scale) of predicted response variance (using response mean as a substitute) and recorded response variance for one neuron for a variational model. Each dot representing one stimulus. 
   {\bf (b)} Distribution of correlations between recorded and predicted response variance for all neurons for the L2+L1, variational-mean (using response mean as a substitute) and variational-variance (calculating response variance), at one random seed. Horizontal lines representing distribution means.
   {\bf (c)} Mean correlations between two response variances (10 seeds per model). Note that variational-variance had lower correlation than the L2+L1. Error bars represent standard deviation of n=10 random seeds for each model. 
   We computed the correlation using the predicted and recorded response variances of the test stimuli for each neuron ($CC=0.34, p=0.002$, Spearman correlation for an exemplary neuron; a). We found that the variational one using response mean as a substitute of variance had a slightly higher mean correlation across neurons compared to the L2+L1 ($p=0.0368$, two-sided permutation tests on 10 random seeds for 10,000 times; b,c).
   }
\label{fig:Fig4_S2}
\end{figure}

\newpage

\subsubsection{Neural activation test with verified MEIs for second dataset} 

\begin{figure}[!htb]
\begin{center}
   \includegraphics[width=\linewidth]{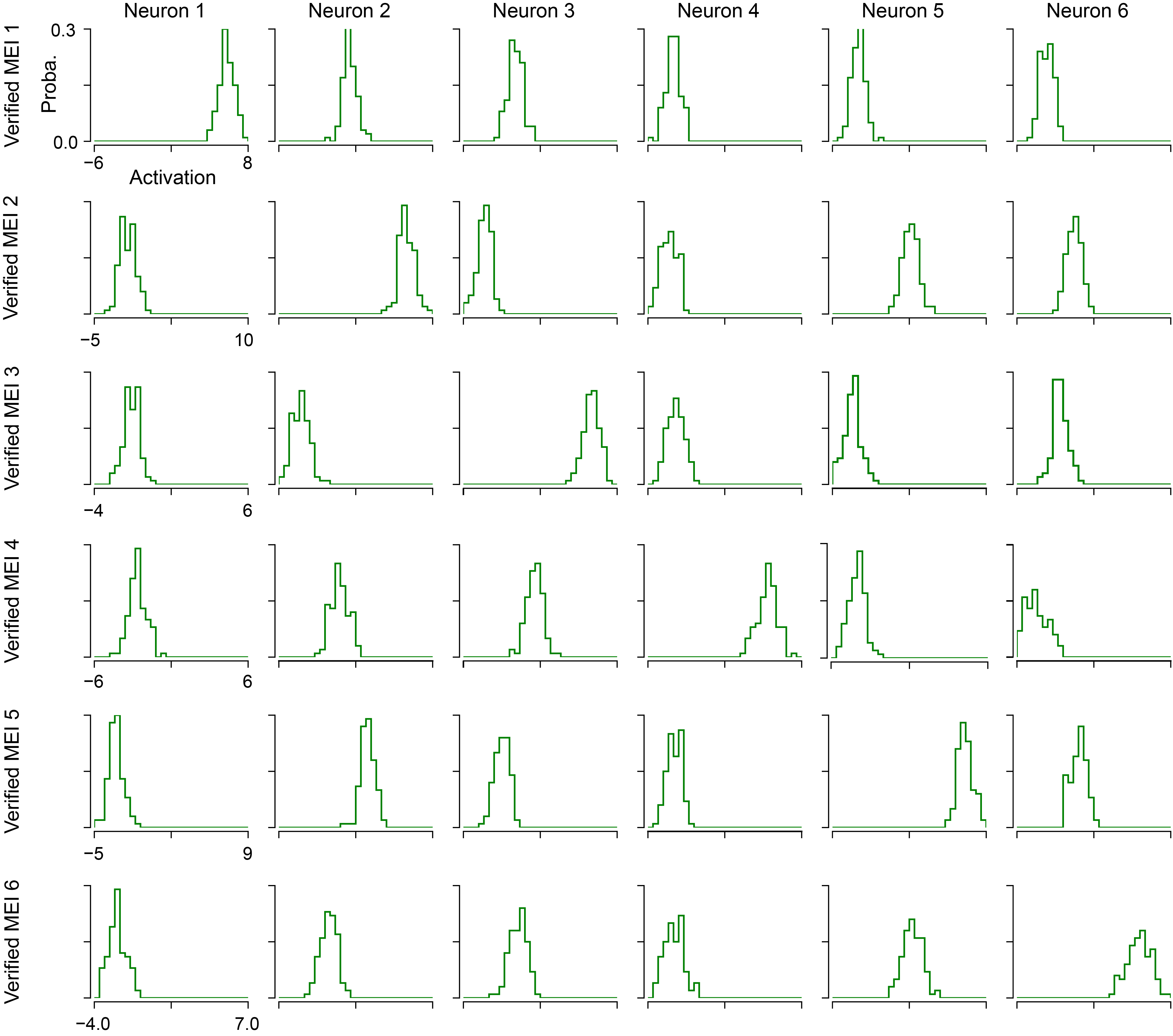}
\end{center}   
   \caption{Neural activation test with verified MEIs for second dataset. 
   Instead of performing closed-loop experiments to examine the effectiveness of our veriational model, we used the verified MEIs from previous study to compare the neuronal activities driven by different stimuli \citep{franke2021behavioral}. We plotted 1D histograms of activation of 6 exemplary cells (from left to right) driven by the verified MEIs (from top to bottom). Interestingly, each MEI from \citet{franke2021behavioral} drove the corresponding neurons optimally.
   }
\label{fig:new_Fig6_S1}
\end{figure}

\end{document}